\documentclass[12pt]{article}

\usepackage{scicite}
\usepackage{lineno}

\usepackage{times}
\usepackage{graphicx}
\usepackage{bm}
\usepackage{color}
\usepackage{mathtools}

\newenvironment{sciabstract}{%
\begin{quote} \bf}
{\end{quote}}

\title{Superconductivity emerging from the N{\'e}el state in {\it infinite-stage} single-layer cuprate La$_2$CuO$_{4+\delta}$}

\author
{Yoshihiko Ihara,$^{1\ast}$ Ramender Kumar,$^{1}$ Kota Miyakoshi$^{1}$, Migaku Oda,$^{1}$ \\
and Kenji Ishida$^{2}$\\
\\
\normalsize{$^{1}$Department of Physics, Faculty of Science, Hokkaido University, Sapporo 060-0810 Japan}\\
\normalsize{$^{2}$Department of Physics, Graduate School of Science, Kyoto University, Kyoto 606-8502 Japan}\\
\\
\normalsize{$^\ast$To whom correspondence should be addressed; E-mail: yihara@phys.sci.hokudai.ac.jp}
}

\date{\today}

\begin{document} 


\baselineskip24pt


\maketitle

%\linenumbers

\begin{sciabstract}
In copper oxides (cuprates) with single CuO$_2$ layer such as La$_{2-x}$Ba(Sr)$_x$CuO$_4$, antiferromagnetism coexists with superconductivity at small doping levels $x$, where chemical disorders are significant. 
Here, we report that superconductivity occurs in a uniform and fully ordered N{\'e}el state in a single-layer cuprate La$_2$CuO$_{4+\delta}$ with a small amount of excess oxygen $(\delta = 0.015)$ 
 as demonstrated by the $^{139}$La nuclear quadrupole resonance measurement. 
A uniform oxygen distribution in the crystal is crucial for achieving microscopic phase coexistence and overcoming the miscibility gap associated with the staging instability; self-organized periodic oxygen arrangement driven by mobile oxygen atoms.  
This finding prompts the reconsideration of superconductivity in cuprates, highlighting that it can emerge in a robust N{\'e}el state that retains sizable magnetic moments and hosts only a small carrier density. 
\end{sciabstract}

\clearpage

{\bf Main Text:}\\ 
%\section*{Introduction}
High transition temperature superconductivity develops in copper oxides  (cuprates) by doping conducting carriers into the Mott-insulating parent material. 
Unpaired electrons of Cu$^{2+}$ ions in the non-doped material are localized due to strong electron interactions, leaving only the spin degrees of freedom active. 
The additional carriers injected into the Mott insulating state drive the frozen charges to itinerant, leading to a significant impact on the conducting properties\cite{Anderson_Science_1987}. 
Superconductivity, which reaches its maximum transition temperature at an optimal doping level of approximately 16 \%, has drawn significant attention, especially since the superconducting (SC) transition temperature $T_c$ exceeds 130 K in Hg-based cuprates \cite{Schilling_Nature_1993}.
The SC mechanism has been explained by the Cooper pair formation through antiferromagnetic (AFM) fluctuations near the AFM criticality \cite{Scalapino_RMP_2012}. 
Before reaching the maximum $T_c$, the lightly doped cuprates exhibit puzzling electronic properties, such as pseudogap \cite{Alloul_PRL_1989}, charge density wave \cite{Wu_Nature_2011,Tranquade_PB_460}, and spin glass states \cite{Cho_PRB_46,Borsa_PRB_52,Keimer_PRB_1992,Wakimoto_PRB_1999}. 
These complex electronic states highlight the diverse physics of cuprates, but hinder emergence of superconductivity in samples with very low doping levels. 

The itinerant carriers are introduced into the Mott-insulating La$_2$CuO$_4$ via chemical substitution, disrupting the periodicity of crystalline lattice. 
A prototypical example is the Sr-substituted La$_{2-x}$Sr$_x$CuO$_4$ \cite{Bednorz_ZPB_1986}, in which Sr$^{2+}$ ions provide conducting holes to the CuO$_2$ layers deforming simultaneously the crystalline lattice by the mismatch in the ionic radius between Sr$^{2+}$ and La$^{3+}$. 
At low doping level, dilute carriers move through variable range hopping, \cite{Oda_SSC_1988} but the structural imperfections cause frequent scattering, leading to carrier localization and a disordered ground state. 
Considerable efforts have been made to preserve a clean electronic state by minimizing the impact on the crystal structure due to doping. 
The SC state on the disorder-free flat CuO$_2$ layer has been revealed in the multi-layered cuprates \cite{Kotegawa_PRB_2004,Mukuda_JPSJ_2012}. 
Although structural disorder in the charge reservoir layers (CRL) causes lattice imperfection in the adjacent outer CuO$_2$ layers (OL), the inner CuO$_2$ layers (IL), shielded by OLs, remain intact and maintain the homogeneous electronic state after doping carriers through OLs. 
Novel electronic properties in the ILs were measured using spectroscopic experiments \cite{Kotegawa_PRB_2004}. 
A striking result from the angle-resolved photoemission spectroscopy (ARPES) measurement demonstrated the microscopic coexistence of superconductivity and antiferromagnetism in the same CuO$_2$ layer. 
Specifically, the SC gap was found on the Fermi surface folded by the periodic AFM spin structure \cite{Kunisada_Science_2020,Kurokawa_NatComm_2023}. 
A pioneering Cu-NMR study estimated the carrier concentrations from the Knight shift measurement and concluded that the SC ground state is already introduced in a sample with only 4 \% of doped carriers when disorders are eliminated \cite{Mukuda_JPSJ_2012}. 
NMR study also detected the coexisting SC and AFM states in the multi-layered cuprates, although the Cu-NMR signals from ILs and OLs were not clearly resolved in the AFM state\cite{Shimizu_PRB_85}.
These findings manifest that superconductivity appears in the AFM state at small doping, posting a question on the pairing mechanism of high-$T_c$ superconductivity.

This intriguing coexistence of phases appearing at doping levels smaller than 8 \% has only been observed in multi-layered cuprates, where the homogeneous ILs are isolated from the CRL. \cite{Mukuda_JPSJ_2012} 
To capture the dawn of high-$T_c$ superconductivity and investigate the noticeable SC properties at low carrier concentrations, a homogeneous CuO$_2$ layer in a single-layer cuprate is needed.
In this study, we focus on a very lightly oxygen doped La$_2$CuO$_{4+\delta}$, in which excess oxygen is distributed in the LaO layers and donates carriers to the CuO$_2$ layer with minimal structural impact. (Fig. 1a) 
At higher concentrations, excess oxygen forms a periodic density modulation along the $c$ direction, disrupting the homogeneous electronic state.
The modulated structure, defined by the number of layers per period $n$, is referred to as stage $n$. \cite{Wells_Science_1997} 
The staging instability restricts oxygen concentrations to integer values of $n$, obscuring much of the phase diagram due to the miscibility gap (Fig.~1b).
The fractional oxygen concentrations are obtained as the mixture of different stages, which creates fractal structure at micrometer length scale. \cite{Fratini_Nature_2010, Poccia_PNAS_2012}   
In the SC La$_2$CuO$_{4+\delta}$ discovered in 1987 \cite{Grant_PRL_1987,Anderson_PRL_1987}, 
the SC fraction of the sample has been identified to have the stage 6 modulation and is isolated from the non-doped La$_2$CuO$_4$ without modulation, which serves only the Mott insulating AFM state \cite{Shirane_PRL_1987, Fratini_Nature_2010, Poccia_PNAS_2012,Jorgensen_PRB_1988,Chaillout_PC_1990,Zolliker_PRB_1990,Zakharov_PC_1994,Wells_Science_1997,Pomjakushin_PC_1997,Hirayama_PRB_1998,Lee_PRB_1999,Savici_PB_2000,Savici_PRB_2002,Imai_PRB_2018}. 
The oxygen concentration at stage 6 reaches $\delta \simeq 0.06$ and is sufficiently high to completely suppress the N{\' e}el state. 
Our sample, synthesized through moderate annealing, \cite{Hundley_PC_1991, Wakimoto_JPSJ_1996} shows a sharp N{\' e}el transition at a temperature reduced from $T_N \simeq 320$ K for non-doped La$_2$CuO$_4$.
At a finite but small $\delta\simeq 0.015$, a homogeneous electronic state is achieved throughout the layers, constructing a distinct infinite-stage phase between the Mott insulating phase at $\delta = 0$ and the first miscibility gap. (Fig. 1b)  
By measuring the infinite stage La$_2$CuO$_{4+\delta}$, we found that even in the single-layer cuprate, superconductivity coexists with AFM state at a microscopic level, as in the case of clean IL in the multi-layered cuprates. 
Surprisingly, the well-known cuprate La$_2$CuO$_{4+\delta}$ allows us to explore the challenging issue of the microscopic coexistence of superconductivity and magnetism. 

X-ray diffraction measurements reveal a $c$-axis length of 13.1453(14) \AA, which corresponds to $\delta \simeq 0.015(1)$ based on the known relationship between $c$ and $\delta$ \cite{Sheptyakov_PC_1999}. 
The doping level was confirmed through magnetization measurements (Supplementary Fig. S1b).
The sharp magnetization peak at the AFM transition and the onset of a strong SC shielding signal (Figs. 1c, 1d, and Supplementary Fig. Sla) determine the AFM and SC transition temperatures as $T_{\rm N} = 266$ K, and $T_c = 32$ K, respectively. 
Superconductivity in the AFM state is microscopically studied by the $^{139}$La-nuclear quadrupole resonance (NQR) experiment, because the strong SC shielding signal does not evidence the microscopic phase coexistence. 
In the paramagnetic state at $T = 295$ K, the spectral width at the NQR frequency of 18.428 MHz ( $m = \pm 7/2 \leftrightarrow \pm 5/2$ transition: Fig. 2a) is only 44 kHz, which is comparable to the spectral width for the non-SC single crystal with $\delta\simeq 0$. \cite{MacLaughlin_PRL_1994}
The unchanged spectral width after oxygen doping corroborates a uniform electric environment near the La site close to the excess oxygen and indicates that the structural periodicity of the CuO$_2$ layers is hardly disrupted. 
The NQR peak splitting was observed below $T_{\rm N}$, because the nuclear-spin energy level is modified by the local magnetic fields generated by the fully ordered Cu$^{2+}$ moments in the N{\'e}el state, as illustrated in Supplementary Fig. S2 \cite{Nishihara_JPSJ_1987}. 
We underline that, in our single-crystal study, the two-peak spectrum in the AFM state shows no trace of paramagnetic signal that should otherwise appear between the peaks. 
The clear two-peak spectrum provides microscopic evidence that the AFM order exists throughout the sample and that the ordered moment is uniform. 
In fact, the spectral width of 20 kHz in the AFM spectra, which is narrower than that above $T_{\rm N}$ and comparable to those for the non-doped single crystal \cite{MacLaughlin_PRL_1994}, indicates that, even with the significant internal fields appearing at the La site, the magnetic broadening by the distribution of the ordered moments is negligibly small in the AFM state. 
Figure 2a shows the temperature variation of the NQR spectra throughout the AFM and SC transition temperatures. 
The peak separation was observed only below $T_N = 266$ K in our sample with finite $\delta$, confirming that the non-SC phase with $T_N>300$ K is absent.
The internal field strength $B_{\rm int}$ extracted from the peak positions at each temperature follows the mean-field type temperature dependence $B_{\rm int} \propto (T_{\rm N}-T)^{\beta}$ with the exponent $\beta = 0.38(3)$. (Fig. 2b) 
The NQR frequency $\nu_Q$ was also obtained from the peak positions and its temperature dependence is shown in Supplementary Fig. S3. 
The significant $B_{\rm int}$ that scales to the one for the non-SC sample with higher $T_N=318$ K \cite{MacLaughlin_PRL_1994} suggests that nearly full magnetic moment is ordered in the infinite-stage phase. 
The SC transition in the N{\'e}el state is indicated by an abrupt drop in the NQR intensity as shown in Fig. 2c. 
The decrease in the NQR intensity is attributed to SC diamagnetism because the SC current screens the radio frequency field within a short distance from the sample surface.
The unchanged spectral shape below $T_c$ confirms that $B_{\rm int}$ remains uniform regardless of the distance from the surface. 
In the remaining NQR signal, still affected by $B_{\rm int}$, the SC transition must be identified to confirm the microscopic coexistence of AFM and SC states.

Exploiting the spectroscopic feature of the NQR measurement, we measured $1/T_1$ using the NQR spectra split by the effect of internal fields and selectively investigated the quasiparticle density of states in the mixed AFM/SC states. 
Details of the measurement procedure are provided in Supplementary Fig. S4. 
Figure 3a shows the temperature dependence of $1/T_1$ measured mainly at high-frequency peaks, 
as well as at low-frequency peaks at several temperatures across $T_c$ to exclude a possibility that one of the peaks overlaps accidentally with the signal from the paramagnetic part of the sample. 
Below $T_N$, $1/T_1$ decreases down to approximately 60 K.
After showing a small hump around 60 K, further decrease in $1/T_1$ was observed below $T_c$ until it turns up to show a broad peak at 5 K. 
The temperature dependence around $T_c$ is highlighted in Fig. 3b by showing $1/T_1T$ as a function of $T$.
Below 60 K, slight increase in $1/T_1T$ was followed by a sudden decrease below $T_c$, which is characteristic of the SC gap opening at the Fermi energy \cite{Imai_JPSJ_1988,Hotta_JPSJ_1993}. 
Figure 3c directly shows the slow recovery of nuclear magnetization in the SC state, without requiring data fitting. 
The observation of the SC gap from the measurement of the NQR spectra affected by the internal field provides clear evidence of the microscopic coexistence of superconductivity and N{\' e}el state. 
The decrease in $1/T_1T$ was interrupted by the low-temperature increase in $1/T_1$, although its height is significantly reduced compared to that for $\delta=0$ (Red circles in Fig. 3a) \cite{Sasaki_JPSJ_1988}. 
Similar low-temperature peak observed in lightly-doped La$_{2-x}$Sr$_x$CuO$_4$ was ascribed to the gradual localization of doped holes \cite{Cho_PRB_46,Chou_PRL_71,Borsa_PRB_52,Niedermayer_PRL_80,Ishida_PRL_92}.
The charge localization does not explain the peak in La$_2$CuO$_{4+\delta}$ because the shift and broadening of NQR spectra were not observed below 5 K.
The remaining fluctuations are more likely related to the glassy ground state in the non-SC La$_2$CuO$_4$ but their origin has not been revealed so far. 
Further speculation on the low-temperature peak is presented in the supplemental material.

The $1/T_1$ measurement using $^{139}$La-NQR spectrum detects the SC gap in lightly oxygen-doped La$_2$CuO$_{4+\delta}$ and reveals the microscopic coexistence of superconductivity and antiferromagnetism in a single-layer cuprate with homogeneous electronic state. 
This finding is further supported by numerical calculations for the Hubbard model \cite{Capone_PRB_2006,Kabayashi_PP_2013}.  
For an oxygen concentration of $\delta \sim 0.015$, the carrier density on a CuO$_2$ layer is estimated at $n = 3.2 \times 10^{20}$ cm$^{-3}$. 
The small $n$ is comparable to the carrier concentration for Li$_x$ZrNCl with $x = 0.1$, \cite{Nakagawa_Science_2021} at which the SC gap crossovers to the pseudo gap associated with the preformed Cooper pairs. 
Surprisingly, even with such small $n$, the energy scale of SC gap $\Delta$ in La$_2$CuO$_{4+\delta}$ is estimated to be approximately 100 K from the power-law temperature dependence of $1/T_1T$ near $T_c$ (dashed line in Fig. 3b, see supplementary information for detail). 
The temperature-dependent $1/T_1T$ above 50 K suggests that the Fermi liquid state is not realized down to 50 K due to the small Fermi energy. 
These findings, along with the large $\Delta$, place La$_2$CuO$_{4+\delta}$ in the intriguing parameter regime of superconductivity, where exotic SC properties have been observed in Li$_x$ZrNCl \cite{Nakagawa_Science_2021} and FeSe. \cite{Kasahara_PNAS_2014} 
Besides the coexisting SC and AFM states, strongly correlated superconductivity in La$_2$CuO$_{4+\delta}$ makes it a crucial material for addressing unresolved questions in cuprate superconductors. 

\clearpage

\includegraphics[width=0.9\linewidth]{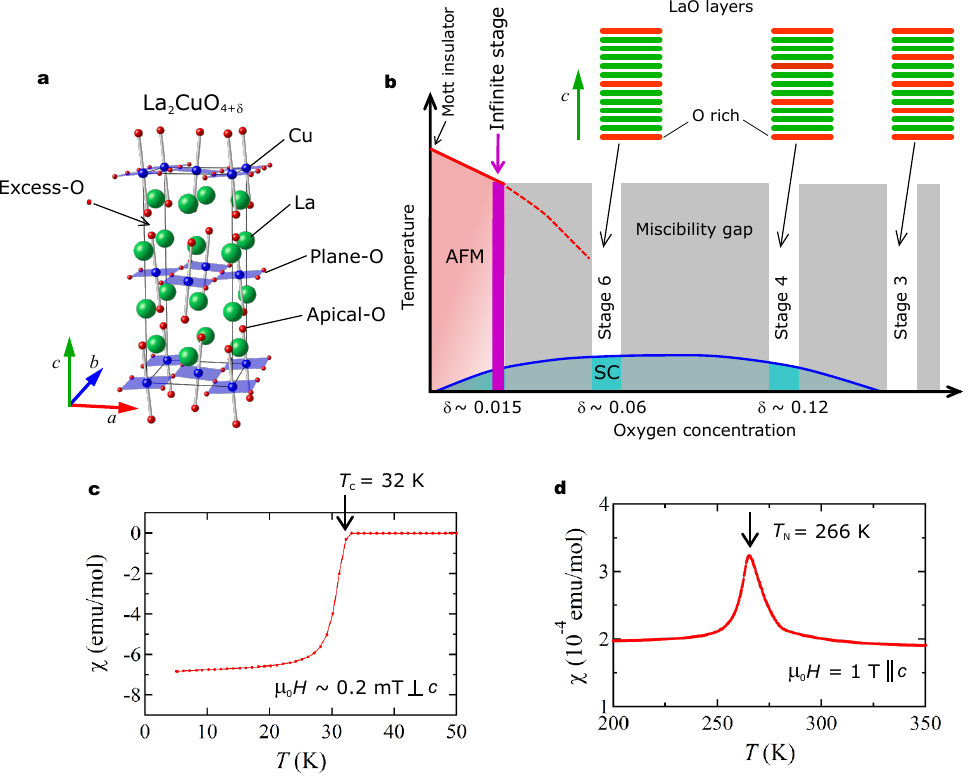}

\noindent {\bf Fig. 1. Superconductivity and antiferromagnetism in the single-layer La$_2$CuO$_{4+\delta}$.} 
{\bf a.} Crystal structure of La$_2$CuO$_{4+\delta}$ with excess oxygen. 
Doped oxygens are diffused in the LaO layers and trapped at interstitial gap. 
The conducting carriers are donated from excess oxygens to the CuO$_2$ layers with minimal structural impact. 
{\bf b.} Schematic phase diagram for La$_2$CuO$_{4+\delta}$ with staging instability and miscibility gaps. 
At each staging phase, a periodic modulation of oxygen density is constructed along the $c$ direction. 
The macroscopic phase segregation is observed in the miscibility gap shaded by gray rectangles. 
The infinite-stage phase with small $\delta \simeq 0.015$, indicated by purple line, has a homogeneous electronic state without the periodic modulation along the $c$ direction. 
{\bf c.} Meissner diamagnetism of the infinite-stage La$_2$CuO$_{4.015}$ below $T_c$. 
The large shielding fraction suggests a sizable superconducting volume fraction. 
The precise estimate of the superconducting volume fraction is, however, difficult due to the uncertainty of large demagnetization factor for the rectangular sample shape. 
{\bf d.} Magnetic susceptibility around the N{\'e}el temperature $T_{\rm N} = 266$ K. The sharp peak at the N{\'e}el transition indicates small distribution of antiferromagnetic transition temperatures in the infinite-stage sample. 
\
\clearpage

\includegraphics[width=0.8\linewidth]{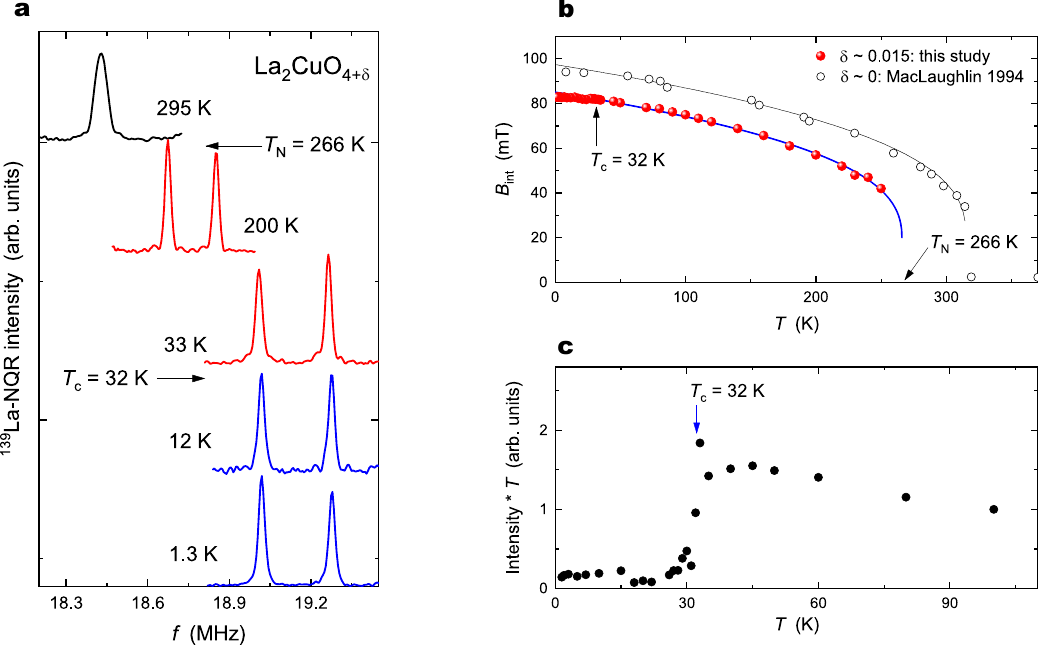}

\noindent {\bf Fig. 2. $^{139}$La-NQR spectra for La$_2$CuO$_{4+\delta}$ in the paramagnetic, antiferromagnetic, and superconducting states. } 
{\bf a.} The spectra for $m = \pm7/2 \leftrightarrow \pm5/2$ transition at representative temperatures across $T_{\rm N}$ and $T_c$. 
In the antiferromagnetic state below $T_{\rm N}$, a uniform internal field splits the NQR spectrum into two sharp peaks. 
No spectrum intensity was observed between the peaks, indicating the absence of distribution in the internal field strength and paramagnetic fraction of the sample.
The peak splitting was observed even in the superconducting state below $T_c$. 
{\bf b.} Temperature dependence of the internal magnetic field $B_{\rm int}$ at the La site calculated from the peak separation. 
The blue solid line represents a mean-field behaviour $(T-T_{\rm N})^{\beta}$ with an exponent $\beta = 0.38(3)$. 
$B_{\rm int}$ for oxygen-doped $\delta \simeq 0.015$ sample scales to the result for the non-SC $\delta \simeq 0$ sample with higher $T_{\rm N} \simeq 318$ K, \cite{MacLaughlin_PRL_1994} confirming the sizable ordered moment. 
{\bf c.} Integrated NQR spectrum intensity multiplied by temperature. An abrupt reduction of the NQR intensity below $T_c$ indicates a substantial increase in the radio frequency shielding effect caused by the SC current. 

\clearpage

\includegraphics[width=0.8\linewidth]{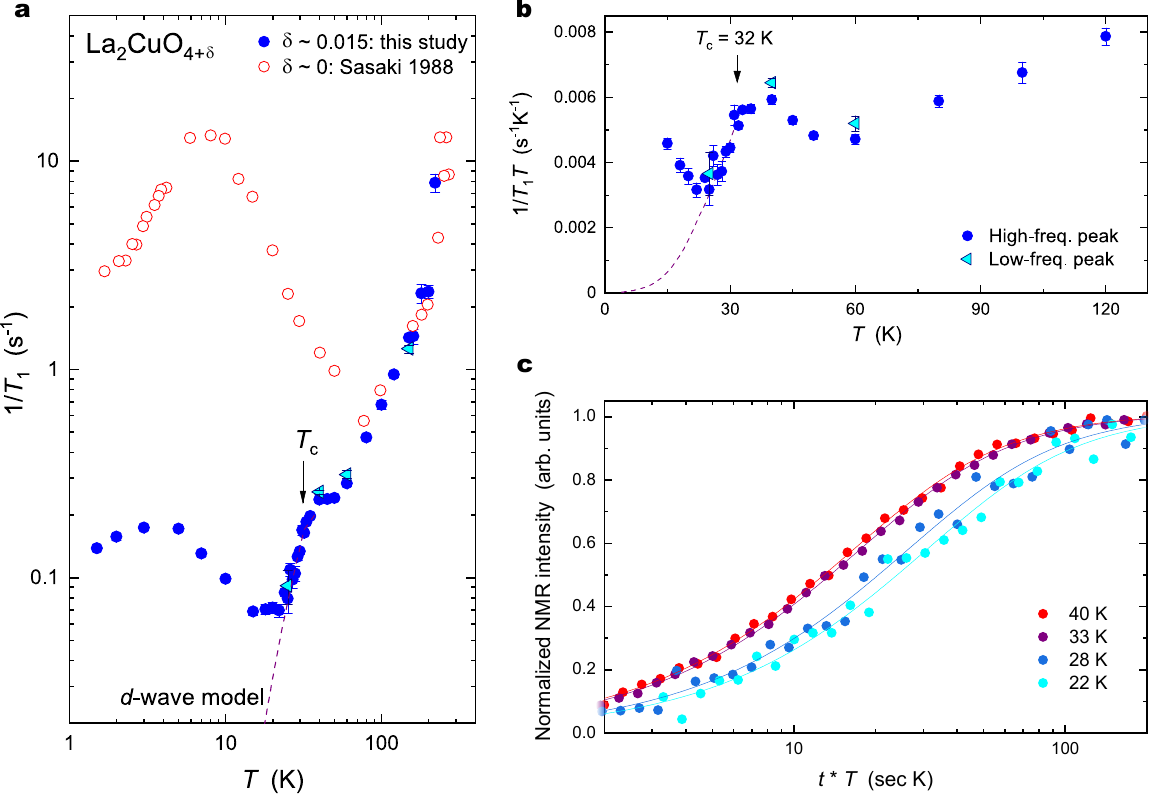}

\noindent {\bf Fig. 3. Superconducting anomaly revealed by the nuclear spin-lattice relaxation rate $1/T_1$ measurement. } 
{\bf a.} The temperature dependence of $1/T_1$ measured with the $^{139}$La-NQR signal affected by $B_{\rm int}$. 
$1/T_1$ was measured both at the high- and low-frequency peaks. 
The consistent results confirm that both NQR peaks originate from the same La site. 
Below $T_{\rm N}$, $1/T_1$ rapidly decreases down to $60$ K.
After showing a small hump, further decrease in $1/T_1$ was observed below $T_c$. 
$1/T_1$ previously reported for the non-SC sample is shown together as the red circles. \cite{Sasaki_JPSJ_1988} 
{\bf b.} The SC anomaly is highlighted by plotting the tempearture dependence of $1/T_1T$ around $T_c$. 
Below 60 K, $1/T_1T$ gradually increases before showing an abrupt decrease at $T_c$. 
The decrease in $1/T_1T$ is ascribed to the appearance of SC gap. 
The simulated temperature dependence for the $d$-wave SC model is represented by dashed lines in {\bf a} and {\bf b}. 
{\bf c.} The recovery profile of NQR intensity after the saturation pulse. 
The horizontal axis is the time interval $t$ after the saturation pulse multiplied by the temperature $T$. 
The reduction of $1/T_1T$ in the SC state is evident without fitting. 

\clearpage

\section*{Methods}

\noindent \underline{Sample preparation}\\
The single-crystalline La$_2$CuO$_{4+\delta}$ sample, grown using the traveling-solvent floating-zone (TSFZ) method shows superconductivity in its as-grown state, though with a broad transition and small volume fraction. 
The single-crystalline rod obtained by the TSFZ method was trimmed into a square shape and the sample was annealed at 900 $^{\circ}$C in Ar atmosphere for $\sim$100 hours to remove excess oxygen included during the TSFZ process and to completely eliminate superconductivity. 
After annealing in Ar, the superconducting (SC) transition completely disappears, and the antiferromagnetic (AFM) transition is observed at $\sim$300 K. 
Subsequently, the deoxidized sample was annealed at 600 $^{\circ}$C in air for $\sim$20 hours. 
The similar annealing condition was reported in previous studies to introduce small amount of oxygen. \cite{Oda_SSC_1988, Hundley_PC_1991, Wakimoto_JPSJ_1996}
The SC volume fraction increases significantly, and a sharp AFM transition is observed at $T_{\rm N} = 266$ K. 
We measured the $c$-axis length to determine the oxygen concentration based on the relationship between $c$ and $\delta$ reported in Ref. \cite{Sheptyakov_PC_1999}. 
The measured $c$-axis length, $c = 13.1453(14)$ \AA~, corresponds to $\delta = 0.015(1)$. 
Assuming that each O$^{2-}$ ion donates two holes to the CuO$_2$ layer, the doping level of our sample is approximately 3 \%. 

\vspace{1em}
\noindent \underline{Magnetization measurement}\\
Magnetization of the single-crystalline sample was measured with SQUID magnetometer (MPMS3, Quantum Design). 
The temperature dependence of the magnetization at 1 T applied along the $c$ direction shows a sharp peak at $T_{\rm N} = 266$ K (Fig. 1d in the main text). 
The core demagnetization of $-0.99 \times 10^{-4}$ emu/mol is subtracted. 
The magnetization peak at $T_N$ is associated with the canted AFM transition due to a finite Dzyaloshinskii-Moriya interaction in addition to the dominant AFM exchange interaction. 
The peak width becomes sharper after annealing. 
Using the relationship between $T_N$ and $\delta$ reported in Ref. \cite{Saylor_PRB_1989}, the oxygen concentration $\delta$ is estimated to be 0.014, consistent with the estimate from the $c$-axis length. 

The SC diamagnetism was measured in the same sample by applying $\sim 0.2$ mT parallel to the CuO$_2$ layer. 
The temperature dependence of magnetization was measured with a zero-field-cooling process:
the sample was cooled to the base temperature in zero magnetic field and the magnetization was measured during the warming after applying a magnetic field at the lowest temperature. 
As shown in Supplementary Fig. S1a, a large Meissner signal was observed below $T_c = 32$ K and the Meissner shielding fraction exceeds -1 at the lowest temperature, confirming the sizable SC volume fraction. 
Precise estimate of the SC fraction is difficult due to the large demagnetization factor associated with the squared sample shape. 

\vspace{1em}
\noindent \underline{Estimate of $\delta$ from spin flop transition}\\
When large magnetic field is applied along the $c$ axis, a jump in magnetization occurs with the flop in canting direction of Cu$^{2+}$ magnetic moments. 
Supplementary Fig. S1b shows the magnetization jump observed for the non-SC sample at 250 K, where the spin flop field is $\sim$3 T. 
We estimated the average size of the ordered moment from the magnetization step $\Delta M$. 
For the non-SC sample before the moderate annealing and SC sample after annealing, we measured the magnetization step as $\Delta M = 3.1 \times 10^{-3}\mu_B/$Cu at 30 K and $2.9 \times 10^{-3} \mu_B/$Cu at 50 K, respectively. 
The decrease in $\Delta M$ is attributed to a reduction in the ordered moment due to carrier doping.
Based on this, we estimate the doping level to be $\sim 0.06 =(3.1-2.9)/3.1$. 
Although this estimate contains a large error, its consistency with the x-ray diffraction results (doping level of $\sim$3 \%) supports the conclustion that excess oxygen introduces a small amount of carriers.  

\vspace{1em}
\noindent \underline{$^{139}$La-NQR measurement and estimate of internal fields}\\
The degenerate energy levels of $^{139}$La nuclear spin $I = 7/2$ are lifted by the interaction between nuclear quadrupole moment and electric field gradient (EFG) (quadrupolar interaction) at zero magnetic fields. 
The energy level scheme with finite quadrupolar interaction is characterized by doubly degenerate four levels, as shown in Supplementary Fig. S2c. 
In the crystal structure of La$_2$CuO$_4$, the EFG at the $^{139}$La site is mainly determined by the ionic charges surrounding the target La site. (Supplementary Fig. S2a) 
As the La site $(0, -0.0088, 0.362)$ is on a mirror plane, one of the EFG principal axes is parallel to the $a$ axis. 
The other two principal axes are within the mirror plane. 
A simple electric field calculation assuming point charges at each ionic site suggests that the largest EFG at the La site is nearly parallel to the La-O bond direction represented by the green bars in Supplementary Fig. S2a. 
In the orthorhombic $Bmab$ structure, the La-O bond is tilted by 7 degrees relative to the $c$ axis.  
If excess oxygen drastically modified the geometry of LaO layers, causing lattice distortion, the $^{139}$La-NQR spectrum would be broadened due to the increase in the distribution of EFG. 
The narrow linewidth of approximately 40 kHz at 18.4 MHz transition confirms that the lattice distortion caused by excess oxygen is negligible. 

In the AFM state below $T_{\rm N}$, internal fields $\bm{B}_{\rm int}$ generated by ordered Cu$^{2+}$ moments further modify the nuclear spin levels through hyperfine interactions, as shown in Supplementary Fig. S2b. 
The level shift depends both on the strength and direction of $\bm{B}_{\rm int}$.
The previous study for $\delta = 0$  \cite{Nishihara_JPSJ_1987,MacLaughlin_PRL_1994} revealed that the internal field strength is approximately 100 mT and direction is 78 degrees from the EFG principal axis. 
The field direction is naturally understood as the vector sum of hyperfine fields from the four near neighbor Cu$^{2+}$ moments in the orthorhombic structure, which yields an internal field perpendicular to the $c$ axis. (Supplementary Fig. S2b) 
We underline here that if only a part of the CuO$_2$ planes conducts while the rest remain in the Mott insulating AFM state, as in staging-type oxygen doping, La sites close to the conducting CuO$_2$ plane would either not experience internal fields or sense weaker fields, since the $^{139}$La nuclear spins couple to only one CuO$_2$ layer. 
The sharply split two-peak structure in the AFM state strongly suggests a uniform internal field, including its orientation relative to the EFG principal axes.  
The internal field strength can be estimated from the peak positions of the $m = +7/2 \leftrightarrow +5/2$ and $-7/2 \leftrightarrow -5/2$ transitions.
These peaks are observed at different frequencies in the AFM state due to the additional Zeeman interaction introduced by the ordered Cu$^{2+}$ moments. 
Assuming that the field orientation remains unchanged in the AFM state, we can independently estimate the internal field strength and the NQR frequency. 
The temperature dependence of the NQR frequency $\nu_Q$, shown in Supplementary Fig. S3, is attributed to thermal contraction of the crystal lattice.

\vspace{1em}
\noindent \underline{Nuclear spin-lattice relaxation rate measurement}\\
The nuclear spin-lattice relaxation rate $1/T_1$ was measured by recording the recovery of $^{139}$La-NQR signal after a saturation pulse. 
The pulse sequence used for the $1/T_1$ measurement is shown in Supplementary Fig. S4a. 
The $^{139}$La-NQR intensity was measured at fixed delays after the saturation pulse. 
The recovery profile, shown for 50 K in Supplementary Fig. S4b, was obtained by repeating the intensity measurement at progressively longer delays. 
The fitting function for the $m = \pm 7/2 \leftrightarrow \pm 5/2$ transition is given as
\begin{equation}
    M(t)=M_0 \left[1-A\left(\frac{3}{14} e^{-3 \frac{t}{T_1} }+ \frac{50}{77} e^{-10 \frac{t}{T_1}}+ \frac{3}{22} e^{-21 \frac{t}{T_1}} \right) \right].
\end{equation}
Here, $M_0$ and $A$ are the nuclear magnetization in the thermal equilibrium and a saturation parameter, respectively. 
The fitted result is shown as a solid line in Supplementary Fig. S4b. 

\clearpage

\section*{Acknowledgments}
The authors would like to acknowledge S. Mizuta and T. Kurosawa for providing high-quality La$_2$CuO$_{4+\delta}$ single crystals grown by the traveling-solvent floating-zone method, Y. Matsushita for providing x-ray diffraction result and  K. Kinjo and S. Kitagawa for supporting NQR experiment.
This work is partially supported by the JSPS Grant-in-Aid for Scientific Research (Grant Nos. 22H04458, 22H00104, 22H00263, 24K06950).

\end{document}


\baselineskip24pt


\maketitle

%\linenumbers

\section*{Contents}
In this supplementary material we provide detailed information about 
\begin{itemize}
    \item Magnetic properties of SC sample with $\delta \simeq 0.015$, (Fig. S1)
    \item Nuclear-spin energy levels with and without the internal fields, (Fig. S2)
    \item Temperature dependence of NQR frequency $\nu_Q$, (Fig. S3)
    \item Nuclear spin-lattice relaxation rate measurement protocol, (Fig. S4)
    \item Peak in $1/T_1$ at low temperature,
    \item Estimate of SC gap size from temperature dependence of $1/T_1$. (Fig. S5) 
\end{itemize} 

\clearpage

\includegraphics[width=0.9\linewidth]{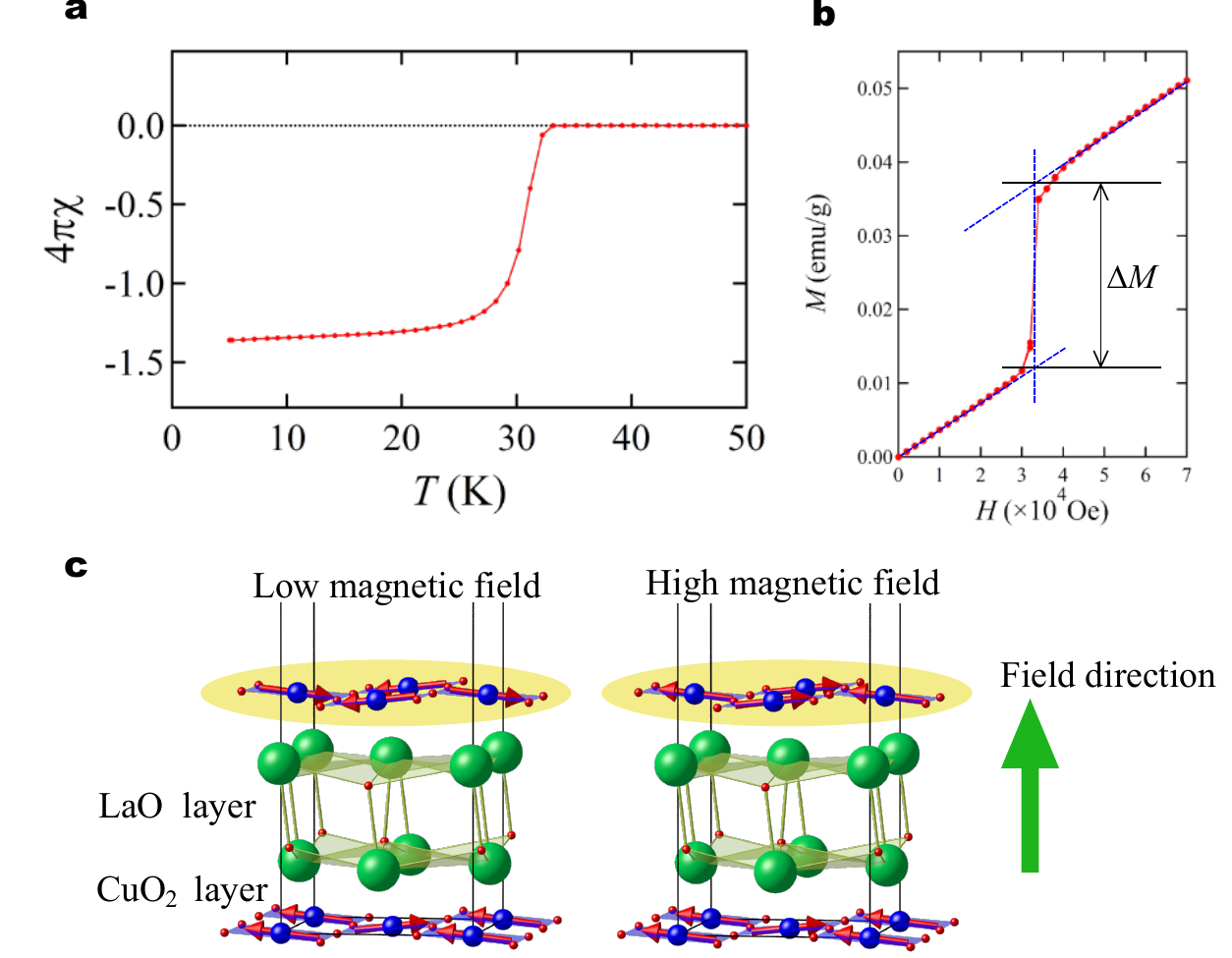}

\noindent {\bf Supplementary Fig. S1. Magnetic properties of SC sample with $\delta\simeq 0.015$.}
{\bf a.} Magnetic susceptibility presented in Fig. 1c is converted to the dimensionless unit. 
At the lowest temperature, the magnetic susceptibility becomes lower than -1 due to the overestimated diamagnetic effect of rectangular sample shape. 
The large shielding fraction suggests the bulk superconductivity, but does not evidence the microscopic coexistence.
Therefore, the phase coexistence should be investigated by a microscopic probe.  
{\bf b.} Magnetization step at the metamagnetic transition. 
The magnetization was measured at 250 K by applying the external fields along the $c$ direction. 
{\bf c.} The magnetic structures above and below the critical fields of $\sim$3 T.
Canted $c$ component of ordered moments are staggered along the $c$ direction in low fields.
High magnetic fields flip magnetic moments in the yellow shade to align the $c$ component of ordered moments along the field direction.
\clearpage

\includegraphics[width=0.8\linewidth]{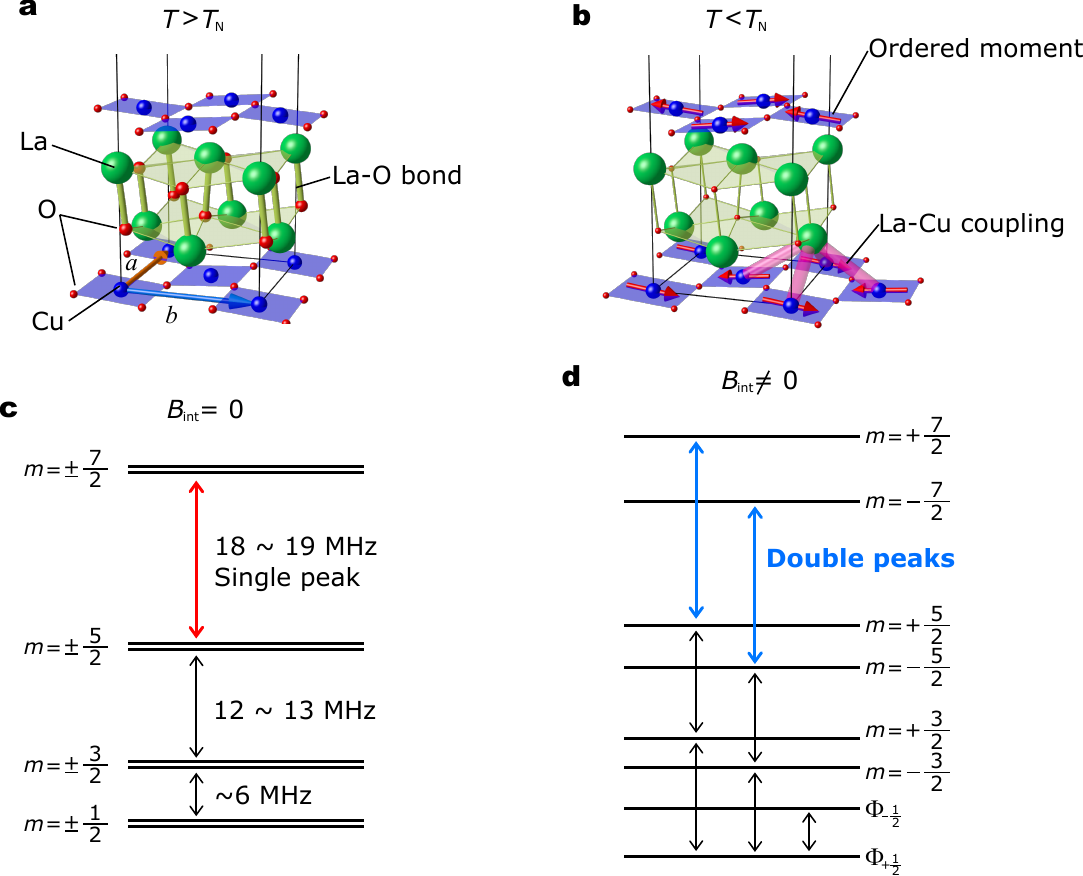}

\noindent {\bf Supplementary Fig. S2. Nuclear-spin energy levels with and without the internal fields.}
{\bf a.} The crystal structure around the NQR target, La sites. 
The EFG at the La site is determined mainly by the oxygen ions surrounding the La site. 
A simple point-charge calculation suggests that the main principal axis of EFG orients along the La-O bond, which is indicated by green rods. 
{\bf b.} The magnetic structure in the AFM state. 
One La nuclear spin is coupled to the neighbouring four Cu$^{2+}$ moments as indicated by thick purple rods. 
{\bf c.}  Energy-level diagram for the La nuclear spin in the paramagnetic state. 
The nuclear levels are lifted only by the electric quadrupole interaction, and thus, the degeneracy between positive and negative magnetic quantum numbers remains. 
In this study, we focus on the highest transition between $m = \pm 7/2 \leftrightarrow \pm 5/2$.  
{\bf d.} Energy-level diagram for the La nuclear spin in the AFM state. 
The internal field generated from the ordered Cu$^{2+}$ moments further splits the degeneracy. 
As a result, the energy separations between $m = +7/2 \leftrightarrow +5/2$ and $m = -7/2 \leftrightarrow -5/2$ are not equal anymore. 
These transitions are thus observed at different frequencies, yielding the peak splitting in the NQR spectrum.

\clearpage
\includegraphics[width=0.8\linewidth]{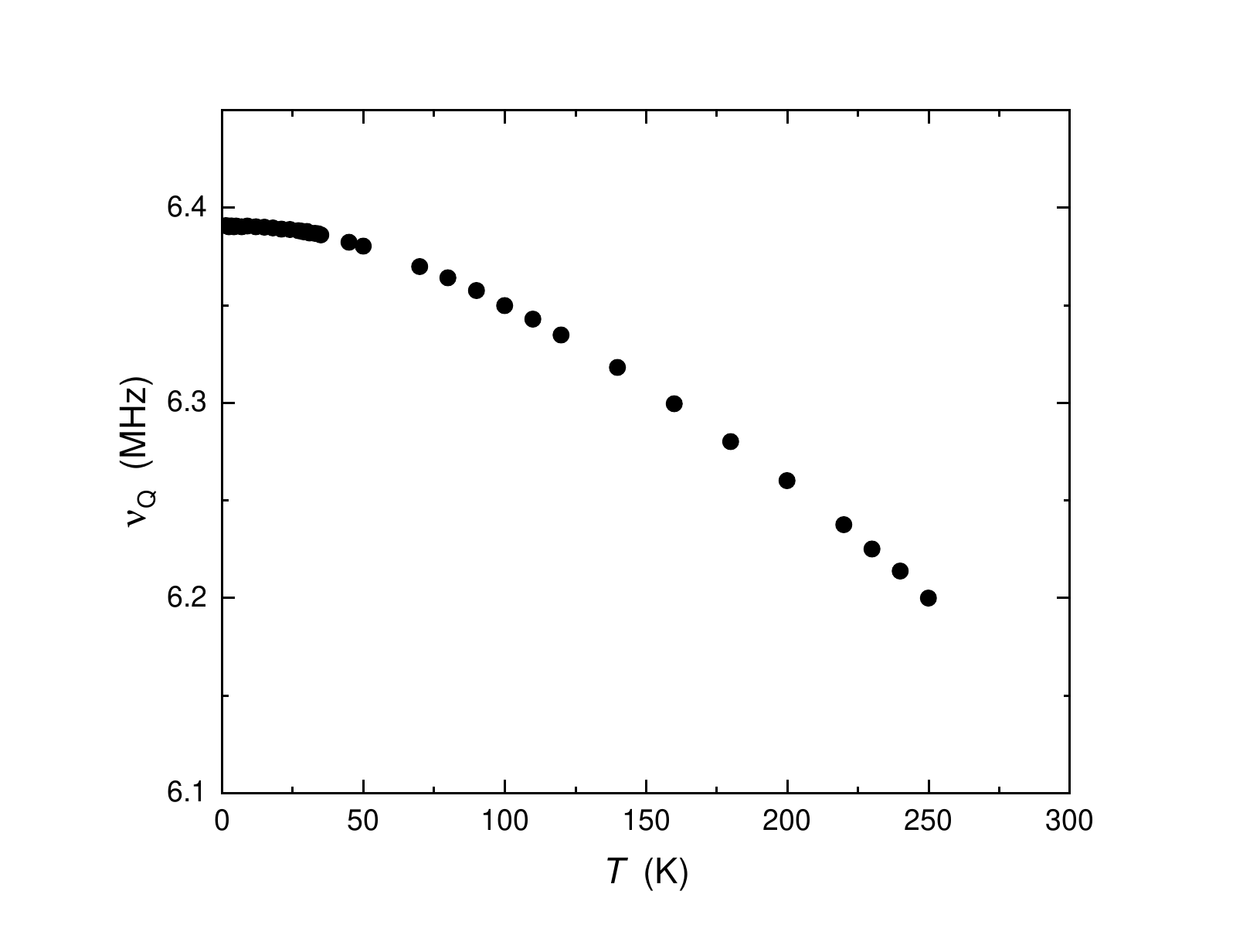}

\noindent {\bf Supplementary Fig. S3. Temperature dependence of NQR frequency $\nu_Q$ extracted from the independent estimate of $\nu_Q$ and $B_{\rm int}$. }
The obtained result is consistent with those reported in the previous study ({\it 37}). 
The gradual increase in $\nu_Q$ toward the lowest temperature indicates no structural anomaly in the AFM state and is simply attributed to the thermal contraction.

\clearpage
\includegraphics[width=0.8\linewidth]{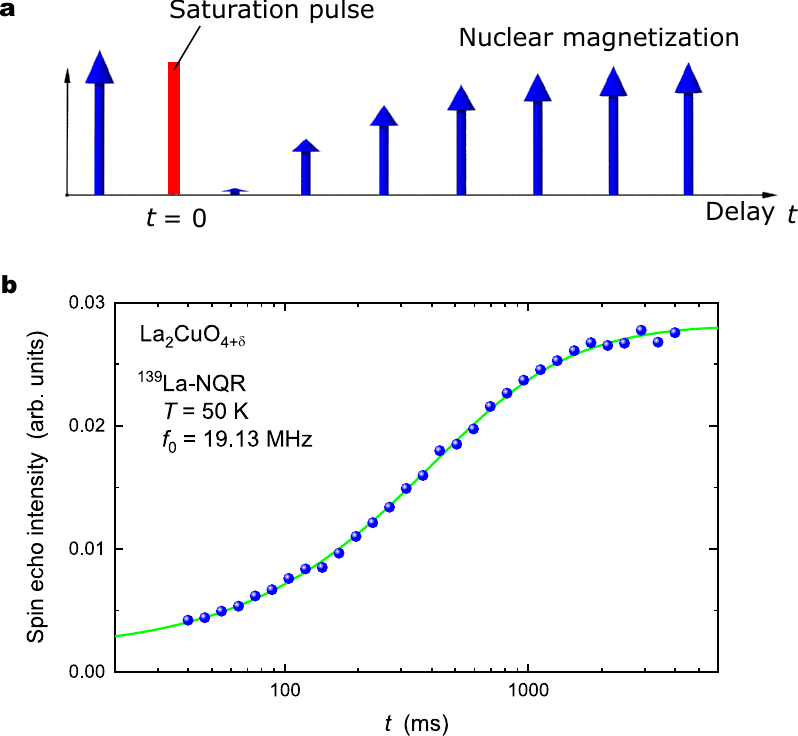}

\noindent {\bf Supplementary Fig. S4. Nuclear spin-lattice relaxation rate measurement protocol.}
{\bf a.} Schematic picture for the recovery of nuclear magnetization. 
We apply the saturation pulse at $t = 0$, which saturates the nuclear magnetization and thus reduces the NQR intensity. 
After the saturation, the nuclear magnetization gradually recovers to the thermal equilibrium value by releasing the energy to the thermal bath through the hyperfine interaction. 
The recovery profile can be measured by repeating the NQR intensity measurements at several delays $t$. 
{\bf b.} A representative nuclear magnetization recovery curve for La$_2$CuO$_{4+\delta}$ measured at 50 K. 
The blue balls are the NQR intensity obtained at a corresponding delay $t$. 
The green solid line is the result of fitting as given in the text. 

\clearpage

\subsection*{Peak in $1/T_1$ at low temperatures}

In the SC state, $1/T_1$ shows an upturn below $T < 0.5T_c$ and a broad peak near 4 K. 
A similar, but more pronounced low-temperature peak was observed in non-SC La$_2$CuO$_4$ without excess oxygen ({\it 41}) and is shown as red circles alongside the present results in Fig. 3A. 
In non-SC La$_2$CuO$_4$, the formation of a glassy magnetic state is suggested by an increase in the $\mu$SR asymmetry signal and absence of a Bragg peak in neutron diffraction studies. ({\it 28}) 
Spin dynamics from the localized part of electronic moments persist in the SC sample, although their amplitude is significantly reduced when conducting carriers are introduced and the SC gap appears at low temperatures. 
Low-energy magnetic fluctuations can persist in the SC state because superconductivity in cuprates uses these magnetic fluctuations as the glue to form Cooper pairs. 
In fact, in the Ni-doped cuprates, the localized Ni$^{2+}$ moments scatter quasiparticles magnetically, leading to an increase in $1/T_1$ due to the dynamics of Ni$^{2+}$ moments in the SC state  \cite{Ishida1993}. 
Furthermore, the localized moments construct a spin glass state through the indirect spin-spin interaction via the conducting quasiparticles without disturbing the SC state \cite{Tokunaga1997}.  

\clearpage

\subsection*{Estimate of SC gap size from temperature dependence of $1/T_1$}
In metallic material, excited nuclear spins relax into thermal equilibrium by transferring energy to conducting electrons through hyperfine coupling.  
Then,  $1/T_1$ is written with hyperfine coupling constant $A_{\rm hf}$ as
\begin{equation} 
    \frac{1}{T_1} =A_{\rm hf}^2 \int N(E)N(E')f(E)\left(1-f(E')\right)\delta(E-E')dEdE'.
\end{equation}
Here, $N(E)$ is the density of states at energy $E$ and $f(E)$ is the Fermi distribution function. 
Conduction electrons at energy $E$ are scattered to $E'$ by absorbing the energy from nuclear spins. 
$\delta(E-E')$ is the delta function to preserve energy conservation. 
In the SC state, the SC gap appearing on the Fermi surface reduces the relaxation rate and the relaxation process is dominated by the quasiparticles thermally excited above the SC gap. 
The temperature dependence of $1/T_1$ in the SC state is, thus, determined by the structure of nodes on the SC gap, from which quasiparticles are generated by small excitation energies. 
The power-law temperature dependence, instead of fully gapped exponential behavior, arises from the anisotropic $d$-wave symmetry of the SC gap. 
The SC gap size $\Delta$ was estimated from the temperature dependence of $1/T_1$ below $T_c$. 
Supplementary Fig. S5 shows the calculated temperature dependence of $1/T_1$ for several gap sizes $R = \Delta /k_B T_c$. 
The experimental data, shown as blue points, lie between the calculated curves for $R = 5$ and $R = 2$, setting the upper and lower bounds for $\Delta$. 
The temperature dependence between 32 K and 20 K is consistently explained by setting $R = 3.5$,
giving an estimate of SC gap $\Delta/k_B \sim$100 K.  
However, by subtracting the contribution from low-temperature peak in $1/T_1$ the SC contribution shows faster temperature dependence and a larger SC gap would be estimated. 

\includegraphics[width=0.8\linewidth]{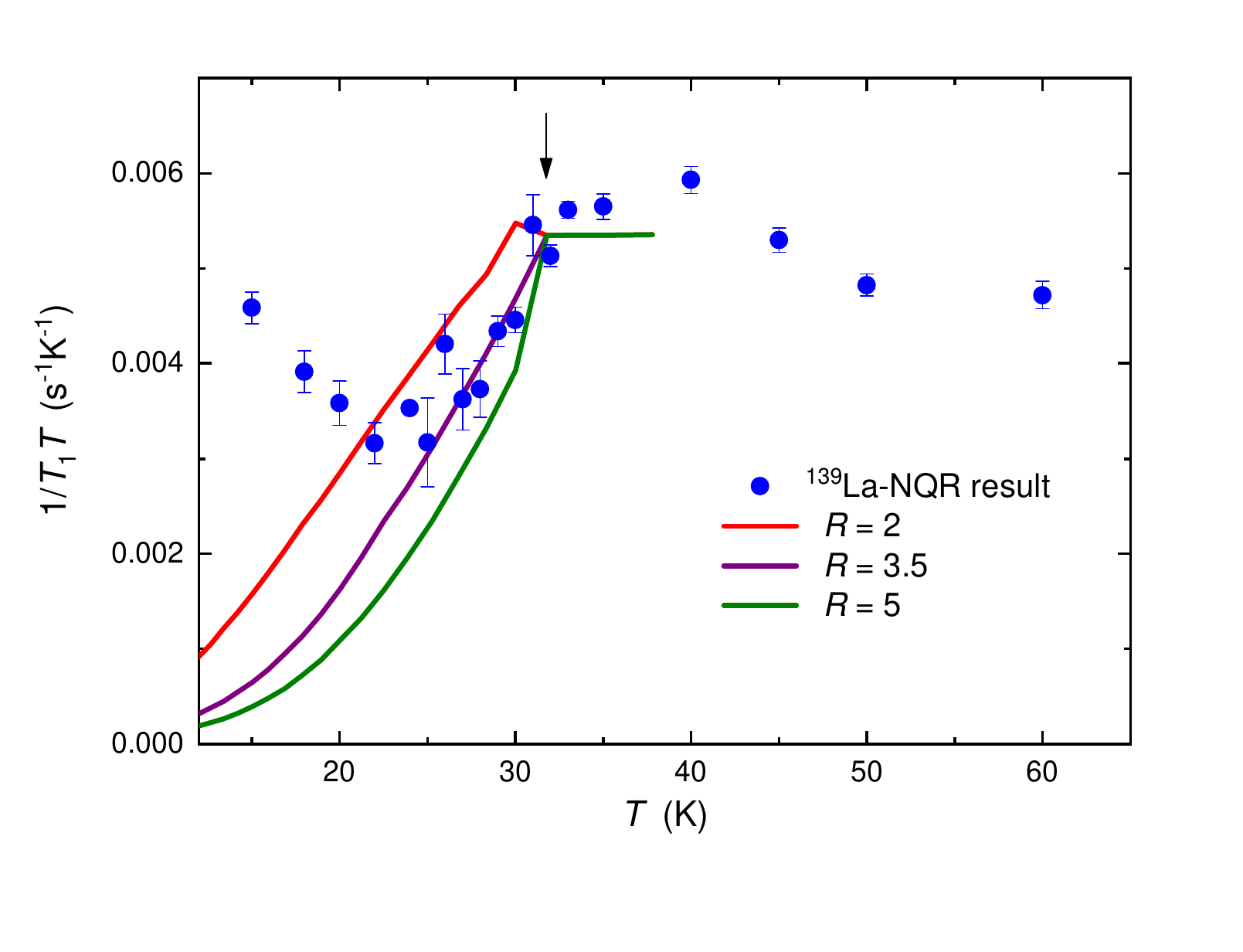}

\noindent {\bf Supplementary Fig. S5. Simulated temperature dependence of $1/T_1T$ for $d$-wave SC model with various gap sizes.} 
The parameter $R$ quantifies the gap size normalized by $T_c$, $R=\Delta/k_BT_c$. 
The experimental results between $T_c$ and 20 K are found between the simulations for $R=2$ (red line) and $R=5$ (green line) and the most suitably explained by setting $R=3.5$ (purple line).